\documentclass[12pt]{iopart}

\usepackage{iopams}

\usepackage{graphicx}

\begin{document}

\title[Temporal Talbot effect for trapped matter waves]{Demonstration of the temporal matter-wave Talbot effect for trapped matter waves}

\author{Manfred J. Mark$^a$, Elmar Haller$^a$, Johann G. Danzl$^a$, Katharina Lauber$^a$, Mattias Gustavsson$^b$ and Hanns-Christoph N\"agerl$^a$}

\address{$^a$ Institut f\"ur Experimentalphysik und Zentrum f\"ur Quantenphysik, \\
Universit\"at Innsbruck, Technikerstra{\ss}e 25, A--6020 Innsbruck, Austria \\
$^b$ Department of Physics, Yale University, P.O. Box 208120, New Haven, Connecticut 06520, USA}
\ead{Christoph.Naegerl@uibk.ac.at}
\begin{abstract}
We demonstrate the temporal Talbot effect for trapped matter waves using ultracold atoms in an  optical lattice. We investigate the phase evolution of an array of essentially non-interacting matter waves and observe matter-wave collapse and revival in the form of a Talbot interference pattern. By using long expansion times, we image momentum space with sub-recoil resolution, allowing us to observe fractional Talbot fringes up to $10^{\rm th}$ order.
\end{abstract}

\vspace{2pc}

\maketitle

\section{Introduction}\label{introduction}

Interference of matter waves is one of the basic ingredients of modern quantum physics. It has proven to be a very rich phenomenon and has found many applications in fundamental physics as well as in metrology \cite{Cronin2009} since the first electron diffraction experiments by Davisson and Germer \cite{Germer1927}. Matter-wave optics has now developed into a thriving subfield of quantum physics. Many key experiments from classical optics have found their counterpart with matter waves, for example the realization of Young's double slit experiment with electrons \cite{Jonsson1961}, the implementation of a Mach-Zehnder-type interferometer with neutrons \cite{Rauch1974}, or, more recently, the observation of Poisson's spot with molecules \cite{Reisinger2009}. The creation of Bose-Einstein condensates (BEC) in 1995 \cite{Anderson1995,Davis1995} opened the door to many more exciting experiments with matter waves, to a large extent in the same way as the laser did in the case of classical light waves.

One remarkable phenomenon in classical optics is the Talbot effect, the self-imaging of a periodic structure in near field diffraction. The effect was first observed by Talbot in 1836 \cite{Talbot1836} and was later explained in the context of wave optics by Rayleigh in 1881 \cite{Rayleigh1881}. When light with a wavelength $\lambda$ illuminates a material grating with period $d$, the intensity pattern of the light passing through the grating reproduces the structure of the grating at distances behind the grating equal to odd multiples of the so-called Talbot length $L_{\rm Talbot}\,{=}\,d^2/\lambda$. At even multiples of the Talbot length the intensity pattern again reproduces the structure of the grating, but shifted laterally in space by half of the grating period. In between these recurrences, at rational fractions $n/m$ of $L_{\rm Talbot}$ (with $n,m$ coprime), patterns with smaller period $d/m$ are formed. This effect is known as the fractional Talbot effect. A necessary requirement for the appearance of the Talbot effect and its fractional variation is the validity of the paraxial approximation \cite{Patorski1989}. Crucial to the Talbot effect is the fact that the accumulated phase differences of the propagating waves behind the grating show a quadratic dependence on lateral distance or grating slit index.

The first observations of the atomic matter-wave Talbot effect \cite{Schmiedmayer1993,Chapman1995} were based on setups comprising an atomic beam and two material gratings, where the second grating acted as a mask used for detection purposes. The demonstration of the fractional Talbot effect with atomic matter waves used the fact that the interference fringes could be recorded directly by using a spatially resolving detector \cite{Nowak1997}. The Talbot effect can also be demonstrated with spatially incoherent wave sources by using an additional first grating to create spatial coherence according to Lau \cite{Lau1948}. In this way, an interferometer is formed that is made of two or even three gratings. Such Talbot-Lau interferometers \cite{Clauser1994} are now an important tool in atomic and molecular interferometry \cite{Cronin2009,Brezger2002,Gerlich2007}. In the context of macroscopic matter waves, i.e. atomic BECs, the Talbot effect has been observed in the time domain by using pulsed phase gratings formed by standing laser waves \cite{Deng1999}. During expansion after release from the trap the BEC was exposed to two short grating pulses separated by a variable time delay and the momentum distribution was measured. At a specific delay, this distribution was observed to rephase to the initial one. In essence, the quadratic dispersion relation of freely propagating, non-interacting matter waves resulted in a quadratic phase evolution for the diffracted momentum states and hence to a temporal version of the Talbot effect. Intriguingly, the Talbot effect is also present for interacting matter waves, as we could show in our previous work \cite{Gustavsson2010}. The momentum distribution of a trapped array of decoupled two-dimensional BECs proved to exhibit a regular, time-varying interference pattern. In this case, the quadratic phase evolution was driven by the local mean-field interaction that had a quadratic spatial dependence reflecting the parabolic shape of the initial density distribution.

In the present work, we report on the demonstration of the temporal Talbot effect using trapped, non-interacting matter waves. Here, the Talbot effect is not driven by interactions but by the (weak) external harmonic dipole-trap confinement, leading to a characteristic quadratic phase evolution. For our measurements we use as before an array of pancake-shaped, two-dimensional BECs in a one-dimensional optical lattice \cite{Gustavsson2010}. The optical lattice takes on the role of the grating. Cancelling the effect of interactions in the vicinity of a Feshbach resonance and decoupling the individual BECs by means of a gravitational tilt initiates long-lived Bloch oscillations (BO) in momentum space \cite{Gustavsson2008}. These are quickly superimposed by a Talbot-type interference pattern in the presence of the external confinement. The pattern can be directly connected to the (fractional) Talbot effect. In particular, after specific hold times that are multiples of the Talbot time, the time-analogue to the Talbot length, a rephasing of the momentum distribution can be observed.

\section{Preparation of the initial sample}\label{preparation}
\begin{figure}
\begin{center}
\includegraphics[width=12cm]{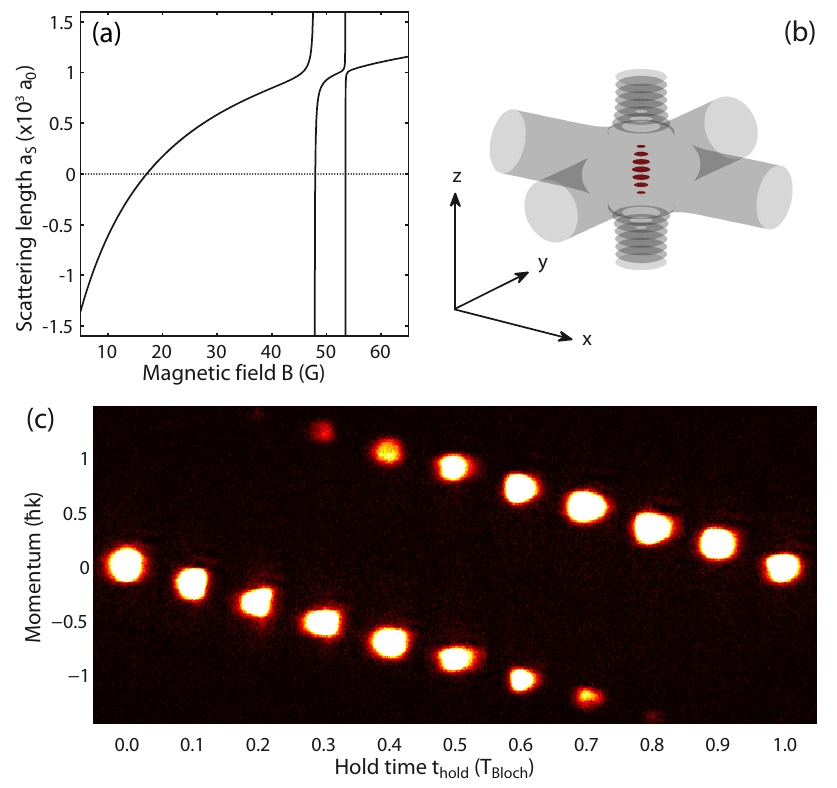}
\caption{\label{figs:fig1} (a) Magnetic-field dependence of the scattering length $a_{\rm s}$ for Cs atoms in $F\,{=}\,3, \ m_F\,{=}\,3$: Wide tunability is given by a broad magnetic Feshbach resonance with a pole near $-11\,$G (not shown), leading to a region with attractive interaction, a zero crossing at about $17\,$G, and a repulsive region above \cite{Chin2004}. Two narrow Feshbach resonances can be seen in the vicinity of $50$ G. (b) Experimental configuration: A vertically-oriented standing laser wave creating a stack of pancake-shaped traps is intersected by two horizontal laser beams. (c) Bloch oscillations: Time series in steps of about $57\,\mu$s showing the quasimomentum distribution over the course of one Bloch cycle.}
\end{center}
\end{figure}

We first produce an essentially pure BEC of Cs atoms (no detectable non-condensed fraction) by largely following the procedure detailed in Ref.~\cite{Weber2003,Kraemer2004}. The atoms are in the lowest hyperfine sublevel $F\,{=}\,3, \ m_F\,{=}\,3$ trapped in a crossed optical dipole trap and initially levitated against gravity by a magnetic gradient field. As usual, $F$ is the atomic angular momentum quantum number, and $m_{\rm F}$ its projection on the magnetic field axis. For the present experiments, the atom number is set to typically $6\times10^4 $ atoms. The trap frequencies in the crossed dipole trap are chosen to be $\omega_{x}\,{=}\,2\pi\times21.7(3)\,$Hz, $\omega_{y}\,{=}\,2\pi\times26.7(3)\,$Hz, and $\omega_{z}\,{=}\,2\pi\times26.9(3)\,$Hz. The confinement along the vertical axis ($z$) and the two horizontal axes ($x,y$) is controlled by two horizontally propagating dipole trap beams with beam waists of $46\,\mu$m and $144\,\mu$m and one vertically propagating dipole trap beam with a beam waist of $123\,\mu$m. The atomic scattering length $a_{\rm s}$ and therefore the strength of interactions in the BEC can be tuned via a magnetic offset field $B$ in a range between $a_{\rm s}\,{=}\,0\,$a$_0$ and $a_{\rm s}\,{=}\,1000\,$a$_0$ by setting $B$ to values between approximately $17$ and $46$ G using a magnetically induced Feshbach resonance \cite{Chin2004} as illustrated in figure\,\ref{figs:fig1}(a). Here, a$_0$ is Bohr's radius. For the initial preparation of the sample, we set $a_{\rm s}$ to positive values, typically between $100\,$a$_0$ and $210\,$a$_0$. Later, $a_{\rm s}$ is set to zero as discussed below. We gently load the condensed atomic sample into a vertical standing wave as illustrated in figure\,\ref{figs:fig1}(b) by exponentially ramping up the power in the standing wave over the course of about $1000\,$ms. The standing wave is generated by a retro-reflected laser beam at a wavelength of $ \lambda\,{=}\,1064.48(5)\,$nm with a $1/e$-waist of about $350\,\mu$m. We are able to achieve well depths of up to $40\,E_{\rm R}$, where $E_{\rm R}\,{=}\,\hbar^2k^2/(2m)\,{=}\,\!h^2/(2 m \lambda^2) \,{=}\, k_{\rm B} \! \times \! 64 \,$nK is the atomic photon recoil energy. Here $k\,{=}\,2\pi/\lambda$, $m$ denotes the mass of the Cs atom, $h$ is Planck's constant, and $k_{\rm B}$ is Boltzmann's constant. The lattice light as well as the light for the dipole trap beams is derived from a single-frequency, narrow-band, highly-stable Nd:YAG laser that seeds a home-built fibre amplifier \cite{Liem2003}. The maximum output power is up to $20\,$W without spectral degradation. The powers in all light beams are controlled by acousto-optical intensity modulators and intensity stabilization servos.

\section{Phase evolution and the Talbot effect}\label{theory}

Our system, the BEC loaded into a 1D optical lattice with spacing $d\,{=}\,\lambda/2$, can be modelled by a discrete nonlinear equation (DNLE) in one dimension \cite{Smerzi2003}, as discussed in our earlier work \cite{Gustavsson2010}. In brief, this equation can be obtained by expanding the condensate wave function from the Gross-Pitaevskii equation, $\Psi$, in a basis of wave functions $\Psi_j(z,r_\perp)$ centred at individual lattice sites with index $j$, $\Psi (z, r_\perp, t)\,{=}\,\sum_j c_j(t) \Psi_j(z,r_\perp)$. Here, $z$ is the coordinate along the (vertical) lattice direction, $ r_\perp$ is the transverse coordinate, and $ c_j(t) $ are time-dependent complex amplitudes. The atoms are restricted to move in the lowest Bloch band and we can write $\Psi_j(r_\perp,z)\,{=}\,w_0^{(j)}(z) \Phi_{\perp}(\rho_j,r_\perp)$, where $w_0^{(j)}(z)$ are the lowest-band Wannier functions localized at the $j$-th site and $\Phi_{\perp}(\rho_j,r_\perp)$ is a radial wave function depending on the peak density $\rho_j$ at each site \cite{Smerzi2003}. By inserting this form into the Gross-Pitaevskii equation and integrating out the radial direction, the DNLE is obtained,
\begin{eqnarray}
    i \hbar \frac{\partial c_j}{\partial t}\,{=}\,J (c_{j-1} + c_{j+1}) + E^{\rm int}_j(c_j) c_j + V_j c_j. \label{eq:1}
\end{eqnarray}
Here, $J/h$ is the tunnelling rate between neighbouring lattice sites, $V_j\,{=}\,\mathfrak{F} d \ j + V_j^{\rm trap}$ describes the combination of a linear potential with force $\mathfrak{F}$ and an external, possibly time-varying trapping potential $V^{\rm trap}_j$, and $ E_j^{\rm int}(c_j)$ is the nonlinear term due to interactions.

We first load the BEC into the vertical lattice and then allow the gravitational force to tilt the lattice potential. We thus enter the limit $\mathfrak{F}d \gg J$, in which tunnelling between sites is inhibited and the on-site occupation numbers $|c_j|^2$ are constant, determined by the initial density distribution. The time evolution of the system is then given by the time-dependent phases of all $c_j$, and the 1D wave function $\tilde \Psi(q,t)$ in quasi-momentum space $q$ acquires a particularly simple form \cite{Witthaut2005}:
\begin{eqnarray}
    \tilde \Psi(q,t) & = & \sum_j c_j(t) e^{-iqjd}\,{=}\,\sum_j c_j(0) e^{-i(\mathfrak{F}dj + V^{\rm trap}_j + E^{\rm int}_{j}) t / \hbar} e^{-iqjd} \nonumber\\
    &=& \sum_j c_j(0) \, e^{-i(q + \frac{\mathfrak{F}t}{\hbar}) jd} \, e^{-i(\beta_{\rm tr} (j-\delta)^2 - \alpha_{\rm int} (j-\delta)^2) t / \hbar} \label{eq:2}
\end{eqnarray}
Here, we have assumed that our external potential is harmonic, given by $V_j^{\rm trap}\,{=}\,\beta_{\rm tr} (j-\delta)^2$, where $\beta_{\rm tr}\,{=}\,m \omega_{\rm z}^2 d^2/2$ characterizes the strength of the potential with trapping frequency $\omega_{\rm z}$ along $z$ for a particle with mass $m$. The parameter $\delta$ in the interval $[-1/2,1/2]$ describes a possible offset of the potential centre with respect to the nearest lattice well minimum along the z-direction. For the interaction term $\alpha_{\rm int}$, the spatial dependence is also parabolic, reflecting the fact that we initially load a (parabolically shaped) BEC in the Thomas-Fermi regime. In our experiments, the offset $\delta$ is not well controlled. It is nearly constant on the timescale of a single experimental run (duration of up to $20$ s), but its value changes over the course of minutes as the positions of the horizontally propagating laser beams generating the trapping potential and the position of the retro-reflecting mirror generating the vertical standing wave drift due to changes of the ambient conditions.

\begin{figure}
\begin{center}
\includegraphics[width=12cm]{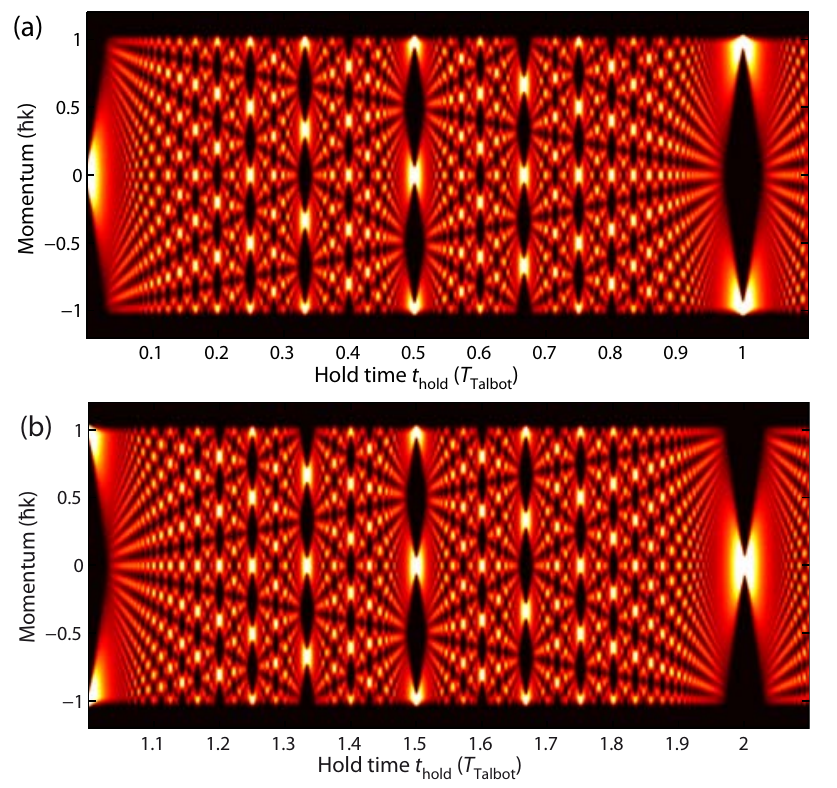}
\caption{\label{figs:fig2} Calculated BEC-based temporal Talbot effect. (a) Momentum distribution as a function of the hold time $t_{\rm hold}$ starting from the initial BEC to the first revival at $T_{\rm Talbot}$ for a pure harmonic potential. White areas indicate a high occupation of the respective momentum state. (b) Same as in (a) from $t_{\rm hold}\,{=}\,T_{\rm Talbot}$ to $t_{\rm hold}\,{=}\,2T_{\rm Talbot}$.}
\end{center}
\end{figure}

The phase evolution in equation \eref{eq:2} has a simple interpretation. The term in the exponent linear in $j$ results in Bloch oscillations \cite{Dahan1996,Anderson1998,Gustavsson2008} with a Bloch period $T_{\rm Bloch}\,{=}\,2\pi\hbar/(\mathfrak{F}d)$. In figure\,\ref{figs:fig1}(c) a full cycle of one BO, corresponding to a Bloch phase from $0$ to $2\pi$, is shown. When restricting ourselves to times that are integer multiples of $T_{\rm Bloch}$ this term can be omitted. The nonlinear exponents proportional to $j^2$ lead to a dephasing between lattice sites, resulting in a time-varying interference pattern for the quasimomentum distribution \cite{Gustavsson2010}. In our experiments we have full control over these nonlinear terms, not only over $\beta_{\rm tr}$ via the external trapping potential, but also over the interaction term characterized by $\alpha_{\rm int}$ via the scattering length $a_{\rm s}$. Our previous work \cite{Gustavsson2010} has focused on the role of interactions, whereas in this work we focus on the (nonlinear) term caused by the external potential. For this we tune $a_{\rm s}$ in such a way that the term with $\alpha_{\rm int}$ is minimized. Now the phase evolution depends only on the term with $\beta_{\rm tr}$. The offset $\delta$ slightly modifies the Bloch period, resulting in a global shift of the interference pattern in quasimomentum space when imaged at integer multiples of the original $T_{\rm Bloch}$. However, as it is irrelevant for the Talbot effect, we set $\delta$ to zero here. By including the simplifications and introducing the Talbot time $T_{\rm Talbot}\,{=}\,h/(m\omega_{\rm z}^2d^2)$, equation \eref{eq:2} reduces to

\begin{eqnarray}
    \tilde \Psi(q,t) & = & \sum_j c_j(0,q) \,e^{-i\pi j^2 t / (2T_{\rm Talbot})}
    \label{eq:3}
\end{eqnarray}

with $ c_j(0,q)\,{=}\,c_j(0)\exp(-iqjd)$. Now the Talbot effect is evident. For times that are even multiples of $T_{\rm Talbot}$ the original wave function is recovered, whereas for odd multiples the original wave function appears with a shift of $\hbar k$ in quasimomentum space. This realisation of the Talbot effect is nearly ideal, since no paraxial approximation is needed and since there is no limitation in time due to decreasing wave packet overlap \cite{Deng1999}. For fractions $n/m$ of $T_{\rm Talbot}$, $m$ copies of the original wave function with a spacing $2\hbar k/m$ appear, corresponding to the fractional Talbot effect. The evolution of the quasimomentum distribution as a function of time can be visualized in terms of so-called matter-wave quantum carpets \cite{Kaplan2000}. Such a quantum carpet, calculated by solving equation \eref{eq:1} numerically with the parameters typical to our experiment, is shown in figure\,\ref{figs:fig2}. Note that in this case the more simple calculation based on equation \eref{eq:3} leads to the same result. However, equation \eref{eq:1} gives us more flexibility in relaxing the requirements of harmonic confinement or negligible tunnelling. We plot the distribution as a line density plot with white areas indicating high densities. Only times that are integer multiples of $T_{\rm Bloch}$ are shown. After a fast spreading of the quasimomentum distribution a regular pattern appears at times for which one expects fractional Talbot interferences. The number of peaks in the momentum distributions directly represents the fraction $t/T_{\rm Talbot}$. At $T_{\rm Talbot}$ a refocusing to the initial distribution occurs, shifted by $\hbar k$ in quasimomentum space. The evolution is then repeated until at $2T_{\rm Talbot}$ the original wave function is recovered.

\section{Experimental realisation}\label{experiment}
For the present experiments we choose a lattice depth of $8\,$E$_R$. For lattice loading the interaction strength is set to $a_{\rm s}\,{=}\,100\,$a$_0$ and the external trap frequencies are changed adiabatically to populate about $40$ lattice sites. After loading, we change $\omega_{z}$ to the final value. This change is done sufficiently quickly (within $3\,$ms) to avoid a change in the initial distribution due to tunnelling, but sufficiently slowly to avoid motional excitations along the $z$-direction. Then, within $0.1\,$ms, we switch off the levitating magnetic field gradient to decouple the individual lattice sites and set the scattering length to the value near $a_{\rm s}\,{=}\,0\,$a$_0$ that gives minimal dephasing \cite{Gustavsson2008}. Note that the point of minimal dephasing does not correspond exactly to $0\,$a$_0$ as residual magnetic dipole-dipole interactions have to be taken into account \cite{Fattori2008}. The shift is calculated to be about $-0.7\,$a$_0$. After a variable hold time $t_{\rm hold}$, which typically corresponds to hundreds of Bloch cycles with $T_{\rm Bloch}\,{=}\,0.575\,$ms, we switch the levitation field back on in $0.1\,$ms and ramp down the optical lattice and the dipole trap responsible for trapping in the vertical direction in $0.3\,$ms. The ramp is adiabatic with respect to the trap frequency of the individual lattice sites, ensuring that the atoms stay in the lowest Bloch band and thus mapping quasimomentum onto real momentum \cite{Kastberg1995}. Before taking an absorption picture we let the sample expand for $80\,$ms while it remains levitated and thus map momentum to real space. The dipole trap responsible for horizontal trapping is not turned off immediately, but instead it is ramped down slowly over the course of $50\,$ms to reduce spreading of the sample in the horizontal direction. At the same time, $a_{\rm s}$ is kept at the value that gives minimal interactions to avoid broadening of the sample in the vertical direction. From the absorption pictures we calculate the momentum width $\Delta p$ as two times the second moment of the momentum distribution along the vertical direction. Note that the presence of the horizontal trap during expansion leads to additional broadening in vertical direction. This broadening plus some residual incoherent background limits the observable values of $\Delta p$. Nevertheless, with our ability to image the quasimomentum with high resolution \cite{Gustavsson2010} we are able to compare not only the momentum width but also the substructure in the momentum distribution to theory.

\begin{figure}
\begin{center}
\includegraphics[width=12cm]{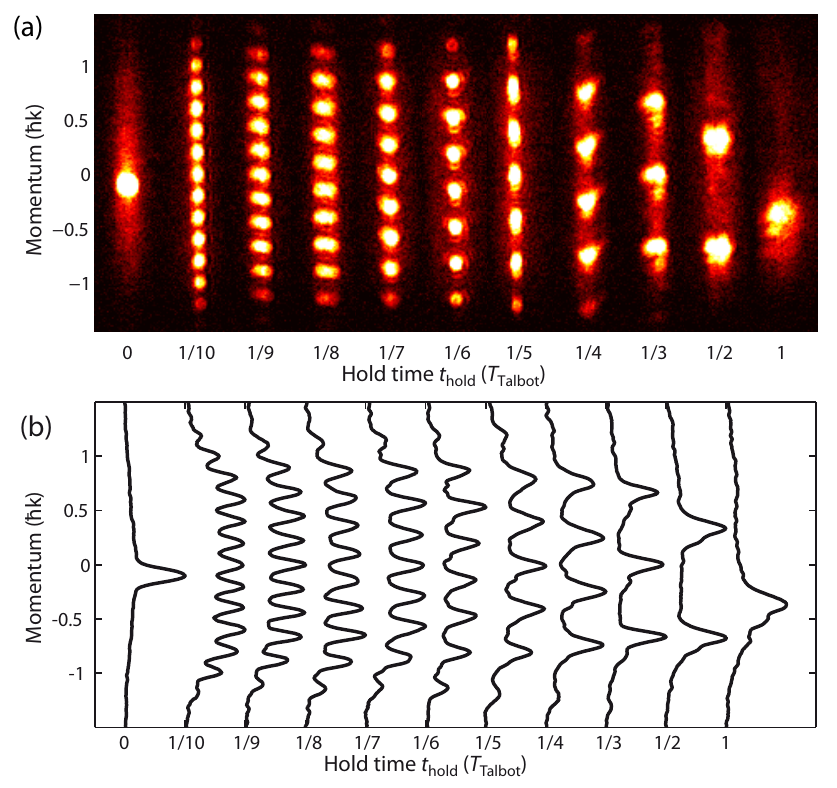}
\caption{\label{figs:fig3} BEC-based temporal Talbot effect - experiment. (a) Series of absorption images after $80$ ms of expansion, showing fractional Talbot fringes of different order in momentum space, starting from the initial momentum distribution of the BEC after two BOs (left), followed by the $10^{\rm th}$ order at $T_{\rm Talbot}/10$, $9^{\rm th}$ order at $T_{\rm Talbot}/9$, etc., down to the $0^{\rm th}$ order at the Talbot time (right). Note that the time axis is not linear. White areas indicate higher density. (b) Horizontally integrated density profiles obtained from the absorption images shown in (a). Note that, e.g. for $T_{\rm Talbot}/10$, the outermost momentum component appears twice, i.e. at both edges of the Brillouin zone. }
\end{center}
\end{figure}

Figure\,\ref{figs:fig3} shows the measured momentum distribution of the atom cloud at specific hold times $t_{\rm hold}$ that are fractions of the calculated Talbot time $T_{\rm Talbot}$. For this measurement we choose a vertical trap frequency of $\omega_{z}\,{=}\,2\pi\times22.0(2)\,$Hz, which gives $T_{\rm Talbot}\,{=}\,555(10)\,$ms. Figure\,\ref{figs:fig3}(a) shows the absorption images as density plots (white areas indicate regions with high density), while figure\,\ref{figs:fig3}(b) plots the horizontally integrated densities from the corresponding images of figure\,\ref{figs:fig3}(a). Initially, the momentum distribution is singly peaked, as expected for a non-dephased BEC. After a rapid coherent dephasing (corresponding to a rapid broadening of the momentum distribution, not shown here) regularly structured patterns appear. The number of peaks within the first Brillouin zone $[-\hbar k, +\hbar k]$ corresponds exactly to the fraction $t_{\rm hold}/T_{\rm Talbot}$, as expected from the theoretical considerations. A small fraction of the atoms is detected outside the first Brillouin zone, likely caused by imperfections in the mapping of quasimomentum onto real momentum. Finally, at the Talbot time, the momentum distribution rephases again to the initial distribution. In general, apart from an overall shift of each individual distribution in quasimomentum space due to variations in $\delta$ as discussed below, we find very good qualitative agreement with the results of the calculation shown in figure\,\ref{figs:fig2}.

\begin{figure}
\begin{center}
\includegraphics[width=12cm]{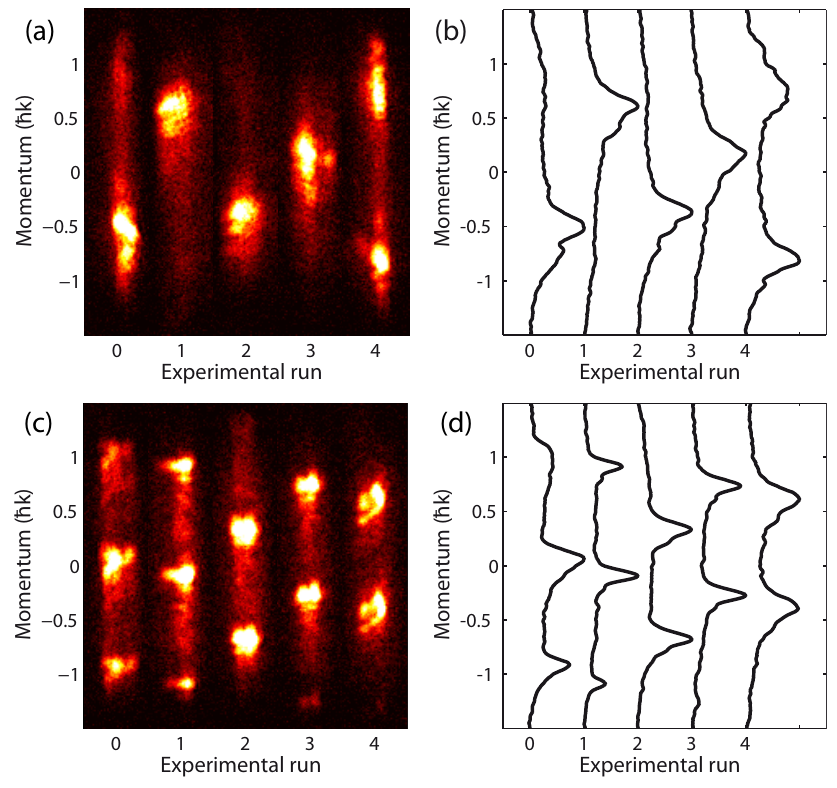}
\caption{\label{figs:fig4} Variations in the momentum distribution between successive experimental realisations for long hold times. (a) Absorption images of five individual experimental realisations with $t_{\rm hold}\,{=}\,T_{\rm Talbot}$. White areas indicate higher density. (b) Horizontally integrated density profiles obtained from the absorption images shown in (a). (c) Absorption images of five individual experimental realisations with $t_{\rm hold}\,{=}\,T_{\rm Talbot}/2$. (d) Horizontally integrated density profiles obtained from the absorption images shown in (c). Note that, in addition to the random shift in quasimomentum space caused by $\delta$, effects of horizontal dynamics, especially fragmentation and density variations along the horizontal axis, can be observed.}
\end{center}
\end{figure}

Figure\,\ref{figs:fig4} illustrates the effect of $\delta$ on the observed patterns in quasimomentum space. For two different hold times $t_{\rm hold}\,{=}\,T_{\rm Talbot}$ and $t_{\rm hold}\,{=}\,T_{\rm Talbot}/2$, absorption images for several individual experimental realisations and the corresponding horizontally integrated densities are shown. The expected single- and double-peaked momentum patterns are reproduced from one experimental realisation to the next, but they experience a varying shift in quasimomentum space. As a consequence of the periodic structure of quasimomentum space, a peak that is located near one edge of the Brillouin zone also reappears at the opposite edge. The maximum possible shift of the pattern in quasimomentum space due to $\delta$ increases with hold time and is calculated to be $\pm\hbar k\times t_{\rm hold}/T_{\rm Talbot}$. This is why the patterns shown in figure\,\ref{figs:fig3}, e.~g. at $t_{\rm hold}\,{=}\,T_{\rm Talbot}$ or at $t_{\rm hold}\,{=}\,T_{\rm Talbot}/2$, agree with the calculated patterns only modulo the shift in quasimomentum space. Note that, alternatively, we could have chosen to present in figure\,\ref{figs:fig3} selected patterns from a sufficiently large sample of measurements, e.g. the one from experimental run 4 for $t_{\rm hold}\,{=}\,T_{\rm Talbot}$ or the one from experimental run 1 for $t_{\rm hold}\,{=}\,T_{\rm Talbot}/2$ shown in figure\,\ref{figs:fig4}.

\begin{figure}
\begin{center}
\includegraphics[width=12cm]{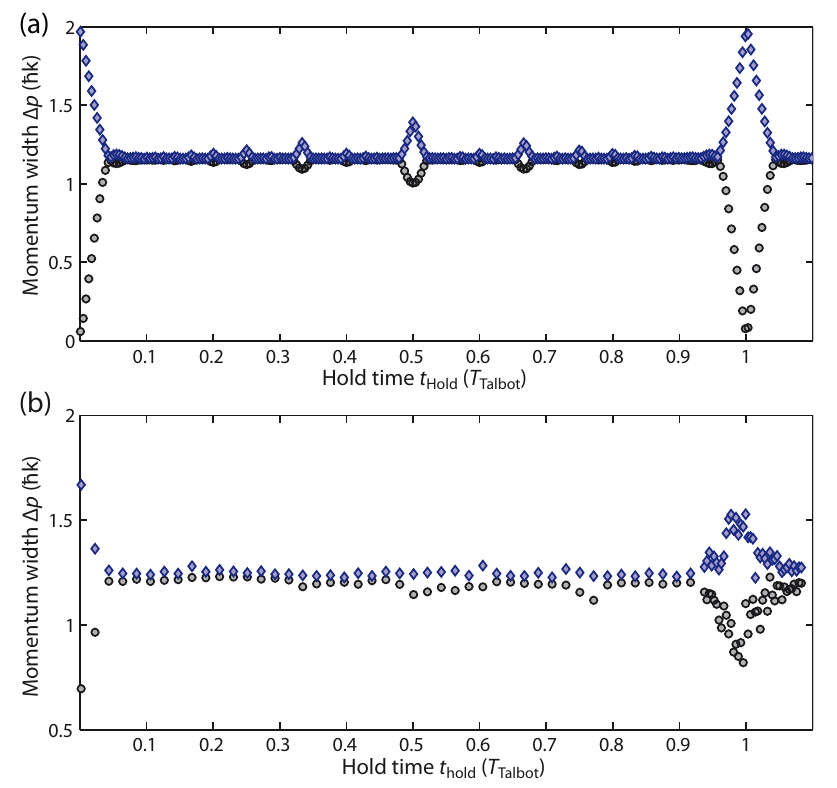}
\caption{\label{figs:fig5} Talbot revival as evidenced by the spread of the momentum width $\Delta p$. (a) Calculated $(\Delta p)_{\rm max}$ (blue diamonds) and $(\Delta p)_{\rm min}$ (black circles) as a function of $t_{\rm hold}$ in units of $T_{\rm Talbot}$. (b) Measurement of $(\Delta p)_{\rm max}$ (blue diamonds) and $(\Delta p)_{\rm min}$ (black circles) as a function of $t_{\rm hold}$ in units of $T_{\rm Talbot}$ for a vertical trap frequency of $\omega_{z}\,{=}\,2\pi\times22.0(2)\,$Hz. The extrema are determined from a sample of $10$ single experimental realisations for each value of $t_{\rm hold}$.}
\end{center}
\end{figure}

A simple quantitative comparison between experiment and calculations can be done by considering the time evolution of the momentum width $\Delta p$. The distribution of this quantity across several experimental realisations is evidently sensitive to the de- and rephasing of the matter wave. In fact, we can relax the choice of the Bloch phase and allow its value to be random. For example, for a non-dephased BEC the momentum width $\Delta p$ is measured to range from $(\Delta p)_{\rm min}\approx0.6\,\hbar k$, corresponding to the singly peaked momentum distribution, to $(\Delta p)_{\rm max}\approx1.7\,\hbar k$ when the momentum distribution is evenly peaked at both edges of the Brillouin zone (e.~g. at half the first Bloch period, see figure\,\ref{figs:fig1}(c)). For a completely dephased sample corresponding to a uniform distribution over the first Brillouin zone we measure a value of $\Delta p\approx1.25\,\hbar k$. Accordingly, the range for the momentum width $\Delta p$ at a given hold time $t_{\rm hold}$ shows a distinct behaviour as a function of $t_{\rm hold}$, in particular indicating the revival at $T_{\rm Talbot}$ by maximizing the difference $D_{\rm \Delta p}\,{=}\,(\Delta p)_{\rm max}-(\Delta p)_{\rm min}$ between the extrema of $\Delta p$. Figure\,\ref{figs:fig5}(a) shows $(\Delta p)_{\rm max}$ and $(\Delta p)_{\rm min}$ as a function of $t_{\rm hold}$ as calculated from equation \eref{eq:1}. Initially and at $T_{\rm Talbot}$ the extrema lie far apart (at these times the calculation gives values for $(\Delta p)_{\rm min}$ that are close to zero in accordance with the fact that the momentum width is determined only by the spread in position space, which is large), whereas at intermediate times the difference is drastically reduced, only increasing slightly at rational fractions of $t_{\rm hold}/T_{\rm Talbot}$. In figure\,\ref{figs:fig5}(b) we plot the measured momentum width extrema. These are determined from samples of $10$ single measurements at each chosen value for $t_{\rm hold}$. The initial rapid collapse agrees well with the fact that the sample dephases. Then, near the calculated value for $T_{\rm Talbot}$, a clear increase in $D_{\rm \Delta p}$ can be seen. The difference recovers almost completely to the initial value. We attribute the slight reduction to additional dephasing mechanisms not included in our simple model as discussed below.

\begin{figure}
\begin{center}
\includegraphics[width=12cm]{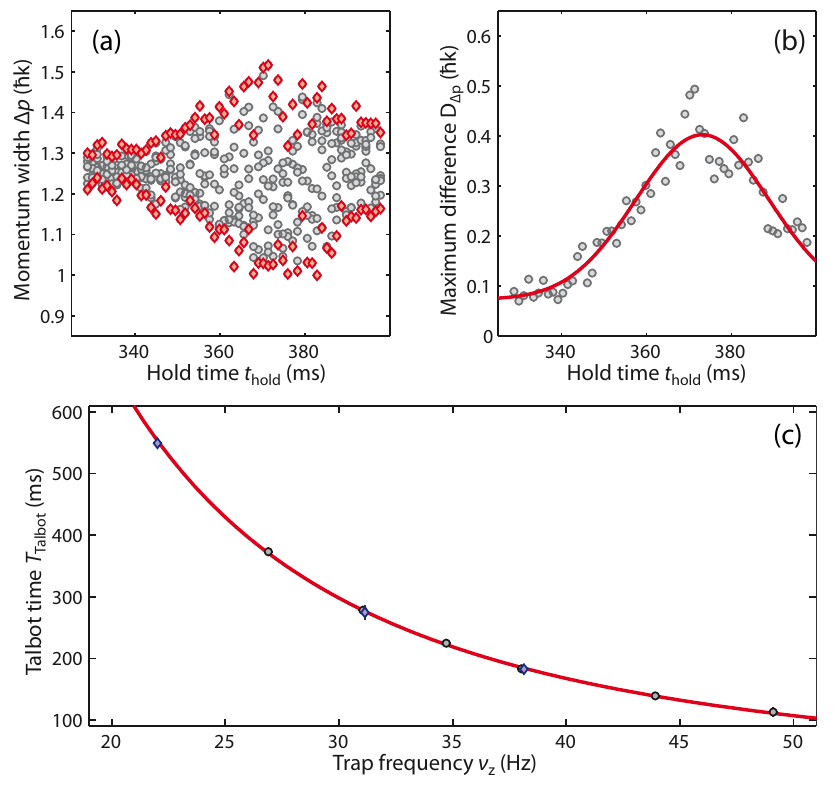}
\caption{\label{figs:fig6} Talbot time $T_{\rm Talbot}$ as a function of the external confinement strength. (a) Momentum width $\Delta p$ in the vicinity of the expected $T_{\rm Talbot}$ for a dipole trap frequency of $\omega_{z}\,{=}\,2\pi\times 26.9(2)\,$Hz for 10 single experimental realisations (black circles). The extrema $(\Delta p)_{\rm max}$ and $(\Delta p)_{\rm min}$ are indicated as red diamonds. (b) Calculated $D_{\rm \Delta p}$ for the measured extrema in (a). The solid line represents a gaussian fit, from which $T_{\rm Talbot}$ is derived. (c) Dependence of $T_{\rm Talbot}$ on the trap frequency $\nu_{z}\,{=}\,\omega_{z}/(2\pi)$. The black (blue) circles (diamonds) represent measurements for which the external harmonic trap is generated by the dipole trap beam with $46\,\mu$m ($144\,\mu$m) beam waist. The solid line gives the calculated values for $T_{\rm Talbot}$. The vertical error bars are the $1 \sigma$ uncertainty of the maximum position of the gaussian fit as shown in (b). The horizontal error bars are equal or smaller than the symbol size.}
\end{center}
\end{figure}

The behaviour of $D_{\rm \Delta p}$ offers a simple method to test the dependence of the Talbot time $T_{\rm Talbot}$ on the vertical trap frequency $\omega_{z}$.  Evidently, $D_{\rm \Delta p}$ has a maximum at $T_{\rm Talbot}$. Figure\,\ref{figs:fig6}(a) shows the momentum width $\Delta p$ in the vicinity of the calculated $T_{\rm Talbot}$, here for a specific trap frequency of $\omega_{z}\,{=}\,2\pi\times 26.9(2)\,$Hz. Again we evaluate 10 experimental realisations for each hold time and select $(\Delta p)_{\rm max}$ and $(\Delta p)_{\rm min}$ to calculate $D_{\rm \Delta p}$. We locate the position of its maximum by a simple gaussian fit, as shown in figure\,\ref{figs:fig6}(b). We then vary $\omega_{z}$ and determine $T_{\rm Talbot}$ accordingly. In Figure\,\ref{figs:fig6}(c) $T_{\rm Talbot}$ is plotted as a function of $\omega_{z}$. The experimental values are in excellent agreement with the calculated values for the Talbot time according to $T_{\rm Talbot}\,{=}\,h/(m\omega_{\rm z}^2d^2)$.

\begin{figure}
\begin{center}
\includegraphics[width=12cm]{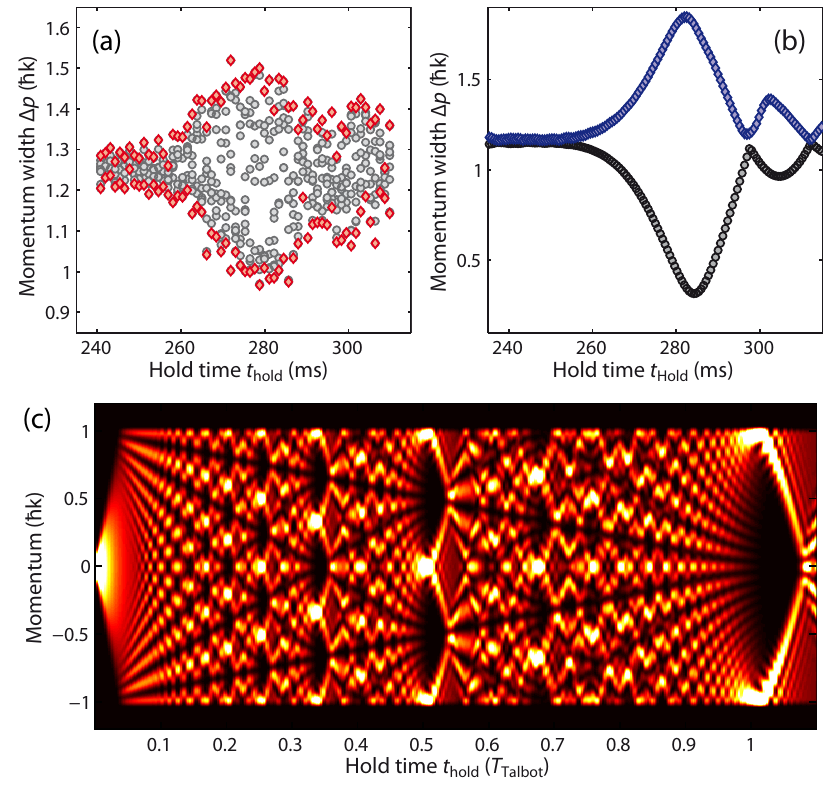}
\caption{\label{figs:fig7} Effect of the anharmonic trapping potential on the momentum distribution. (a) Momentum width $\Delta p$ in the vicinity of the expected $T_{\rm Talbot}$ for a dipole trap frequency of $2\pi\times31.1(2)\,$Hz for 10 single experimental realisations (black circles). The vertical dipole trap is created by the more tightly focused dipole trap beam with a beam waist of $46\,\mu$m. The extrema $(\Delta p)_{\rm max}$ and $(\Delta p)_{\rm min}$ are indicated as red diamonds. (b) Calculated $(\Delta p)_{\rm max}$ and $(\Delta p)_{\rm min}$  in the vicinity of the expected $T_{\rm Talbot}$ for the same experimental parameters as in (a). For the trapping potential the real gaussian shape of the dipole trap is used. (c) Full calculation of the momentum distribution as a function of the hold time $t_{\rm hold}$ using the same parameters as in (b).}
\end{center}
\end{figure}

We finally discuss the main limitations for our experiment. We believe that the total number of subsequent revivals that we can observe (we detect up to 4 revivals) is mainly limited by three-body loss and by the anharmonicity of the trapping potential. Three-body loss heats the two-dimensional BECs residing at each lattice site. This leads to a loss of phase coherence and thus decreases the visibility of the revivals. Perhaps more interestingly, the anharmonicity of the trapping potential along the vertical direction causes deviations from the quadratic phase evolution required for the Talbot effect. In order to test this effect we generate the vertical trapping potential with a more tightly focused dipole trap beam, which enhances the effect of anharmonicity. We then observe non-perfect Talbot revivals followed by subrevivals as can be seen in figure\,\ref{figs:fig7}(a). This is in qualitative agreement with calculations shown in figure\,\ref{figs:fig7}(b), for which the real gaussian shape of the trapping potential instead of a simple harmonic one has been used. The full calculated time evolution of the momentum distribution is shown in figure\,\ref{figs:fig7}(c). The distortion of the matter-wave quantum carpet can clearly be seen.

\section{Conclusion}\label{conclusion}

We have demonstrated the temporal Talbot effect with trapped, non-interacting matter waves. High resolution imaging in quasimomentum space allows us to resolve Talbot fringes up to the 10th order. We have tested the dependence of the Talbot time on the strength of the confinement and have found very good agreement with the calculated value. We find that the interference pattern is sensitive to the anharmonicity of the trapping potential. In principle, the detailed structure of the interference pattern and the precise revival times are sensitive probes for force gradients and interactions between atoms. The weak magnetic dipole-dipole interaction, for example, has recently been investigated in the context of matter-wave interferometry \cite{Fattori2008}. Matter-wave interferometry in the Talbot regime could potentially be used to examine in detail the effect of the long-range nature of such an interaction. 
Similarly, a spatially dependent force like the Casimir-Polder force \cite{Casimir1948,Harber2005,Chwedenczuk2010} near a surface could be investigated through its influence on the Talbot interference pattern.

\ack

We are indebted to R. Grimm for generous support and we thank A. Daley for valuable discussions. We gratefully acknowledge funding by the Austrian Science Fund (FWF) within project I153-N16 and within the framework of the European Science Foundation (ESF) EuroQUASAR collective research project QuDeGPM.

\clearpage

\clearpage

\clearpage

\clearpage

\clearpage

\clearpage

\clearpage

\clearpage

\section*{References}


\end{document}